\documentclass[preprint,showpacs,preprintnumbers,amsmath,amssymb]{revtex4}

\usepackage{graphicx}
\usepackage{dcolumn}
\usepackage{bm}

\begin{document}

\preprint{APS/123-QED}

\title{$\beta$-decay study of $^{77}$Cu}

\author{N. Patronis}

\altaffiliation[Present address: ]{%
Department of Physics, The University of Ioannina, GR-45110 Ioannina, Greece}%

 \email{npatronis@cc.uoi.gr}

\affiliation{Instituut voor Kern- en Stralingsfysica, K.U.Leuven, Celestijnenlaan 200D,
B-3001 Leuven, Belgium}%

\author{H. De Witte}
\affiliation{Instituut voor Kern- en Stralingsfysica, K.U.Leuven, Celestijnenlaan 200D,
B-3001 Leuven, Belgium}%
\author{M. Gorska}
\altaffiliation[Present address: ]{%
GSI, Planckstrasse 1, D-64291 Darmstadt, Germany }%
\affiliation{Instituut voor Kern- en Stralingsfysica, K.U.Leuven, Celestijnenlaan 200D,
B-3001 Leuven, Belgium}%
\author{M. Huyse}
\affiliation{Instituut voor Kern- en Stralingsfysica, K.U.Leuven, Celestijnenlaan 200D,
B-3001 Leuven, Belgium}%
\author{K. Kruglov}
\affiliation{Instituut voor Kern- en Stralingsfysica, K.U.Leuven, Celestijnenlaan 200D,
B-3001 Leuven, Belgium}%
\author{D. Pauwels}
\affiliation{Instituut voor Kern- en Stralingsfysica, K.U.Leuven, Celestijnenlaan 200D,
B-3001 Leuven, Belgium}%
\author{K. Van de Vel}
\affiliation{Instituut voor Kern- en Stralingsfysica, K.U.Leuven, Celestijnenlaan 200D,
B-3001 Leuven, Belgium}%
\author{P. Van Duppen}
\affiliation{Instituut voor Kern- en Stralingsfysica, K.U.Leuven, Celestijnenlaan 200D,
B-3001 Leuven, Belgium}%
\author{J. Van Roosbroeck}
\affiliation{Instituut voor Kern- en Stralingsfysica, K.U.Leuven, Celestijnenlaan 200D,
B-3001 Leuven, Belgium}%

\author{J.-C. Thomas}
\affiliation{Grand Accelarateur National d'Ions Lourds (GANIL),
CEA/DSM - CNRS/IN2P3, B.P. 55027, F-14076 Caen Cedex 5, France
}%

\author{S. Franchoo}
\altaffiliation[Present address: ]{%
Institut de Physique Nucl\'{e}aire, IN2P3-CNRS/Universit\'{e} Paris-Sud, F-91406 Orsay Cedex, France }%
\affiliation{ISOLDE, CERN, CH-1211, Gen{\`e}ve 23, Switzerland
}%
\author{J. Cederkall}
\affiliation{ISOLDE, CERN, CH-1211, Gen{\`e}ve 23, Switzerland
}%
\author{V.N. Fedoseyev}
\affiliation{ISOLDE, CERN, CH-1211, Gen{\`e}ve 23, Switzerland
}%
\author{H. Fynbo}
\altaffiliation[Present address: ]{%
Department of Physics and Astronomy, University of \r{A}rhus, DK-8000 \r{A}rhus, Denmark}
\affiliation{ISOLDE, CERN, CH-1211, Gen{\`e}ve 23, Switzerland
}%
\author{U. Georg}
\affiliation{ISOLDE, CERN, CH-1211, Gen{\`e}ve 23, Switzerland
}%
\author{O. Jonsson}
\affiliation{ISOLDE, CERN, CH-1211, Gen{\`e}ve 23, Switzerland
}%
\author{U. K{\"o}ster}
\altaffiliation[Present address: ]{%
Institut Laue Langevin, 6 rue Jules Horowitz, F-38042 Grenoble Cedex 9, France}
\affiliation{ISOLDE, CERN, CH-1211, Gen{\`e}ve 23, Switzerland
}%

\author{T. Materna}
\affiliation{%
Institut Laue Langevin, 6 rue Jules Horowitz, F-38042 Grenoble Cedex 9, France}

\author{L. Mathieu and O. Serot}
\affiliation{%
CEA-Cadarache, DEN/DER/SPRC/LEPh, F-13108 Saint
Paul lez-Durance, France}

\author{L. Weissman}
\affiliation{National Supercontacting Cyclotron Laboratory, Michigan State University, 164 S. Shaw Lane, Michigan 48824-1312, USA
}%
\author{W.F. Mueller}
\affiliation{National Supercontacting Cyclotron Laboratory, Michigan State University, 164 S. Shaw Lane, Michigan 48824-1312, USA
}%

\author{V.I. Mishin}
\affiliation{Institute of Spectroscopy, Russian Academy of Sciences, 142092 Troitsk, Russia
}%

\author{D. Fedorov}
\affiliation{St. Petersburg Nuclear Physics Institute, 188350 Gatchina, Russia
}%


\date{\today}

\begin{abstract}
A $\beta$-decay study of $^{77}$Cu has been performed at the
ISOLDE mass separator with the aim to deduce its $\beta$-decay properties 
and to obtain spectroscopic information on $^{77}$Zn.
Neutron-rich copper isotopes were produced by means of proton- or neutron-induced fission reactions on 
$^{238}$U. After the production,  $^{77}$Cu was selectively laser ionized,
mass separated and sent to different detection systems where $\beta-\gamma$ and $\beta-n$  coincidence data were collected. We report on the deduced half-live,
decay scheme, and possible spin assignment of $^{77}$Cu. 
\end{abstract}

\pacs{23.20.Lv; 23.40-s; 27.50+e; 21.10.Tg}
\maketitle

\section{Introduction} \label{intro}

Recently, theoretical as well as experimental studies indicate that the size of shell gaps can
 alter when changing the N/Z ratio leading to changes in magic numbers when going away from the
 valley of stability \cite{sor08}. An interesting region of the nuclear chart to study this 
phenomenon is situated around $^{68}$Ni because of the closed proton
 shell (Z=28) and the closed harmonic-oscillator sub-shell (N=40). 
The robustness of the Z=28 shell closure as neutrons are added in the unique-parity orbit 
$\nu$g$_{9/2}$ is of special interest and the induced core-polarization effects
have been studied together with the fragile nature of the N=40 sub-shell closure by means of different experimental approaches. Coulomb-excitation studies in the neutron rich
Ni isotopes \cite{bre08,sor02} resulted in small B(E2) values for
 N=40  suggesting a new sub-shell closure in concert with the high excitation energy of the 2$^{+}_1$ state \cite{bro95}. On the other hand, Grawe {\it et al.} \cite{gra01} and Langanke {\it et al.} \cite{lan03} explained the small observed B(E2) strength and the high excitation energy of the 2$^{+}_1$ state in a different way. On the same ground, recent mass measurements in this region show a very weak irregularity in the two-neutron-separation energies around N=40 giving no evidence for a well-established shell closure \cite{gue07,rah07}. The same mass region was further studied by means of Coulomb excitation and $\beta$-decay experiments in neighboring Cu and Zn neutron rich isotopes \cite{stef08,stef07,van07,van09,mue99,fra98,van05}. From those studies the fragile nature of the N=40 sub-shell closure was confirmed and the key role of the g$_{9/2}$ orbit was revealed in the systematics. A common conclusion in the above mentioned surveys is the need for further spectroscopic information in this region and especially towards the more exotic nuclei around the doubly magic $^{78}$Ni. 

Another interest for the region around Z=28 and 40$<$N$<$50 is related to the astrophysical rapid neutron-capture nucleosynthesis process i.e. the r-process. The r-process is responsible for the origin of about half of the elements heavier than iron. Nuclei with closed neutron shells represent a special set of waiting points which cause a vertical diversion of the r-process path in the neutron-rich side of the chart of nuclei. $^{78}$Ni is a doubly-magic nucleus that represents the most important waiting point of the r-process \cite{kra93}. For this reason the $\beta$-decay properties of the nuclei in the region as well as the spectroscopic information is of paramount importance in r-process calculations.   

In this paper we report on the $\beta$-decay study of $^{77}$Cu. This study is part of a systematic $\beta$-decay study at ISOLDE, CERN \cite{kug00} of neutron-rich Cu isotopes ranging from $^{71}$Cu to $^{78}$Cu. The odd-odd cases have been described in \cite{stef07,van204,van04,van02,tho06}. The results on the odd-mass $^{71-75}$Cu are discussed in the PhD work of T. Faul \cite{fau08}. The lowering of the first-excited states with spin and parities 5/2$^{-}$ and 1/2$^{-}$, as the neutron number increases in odd copper isotopes with A$>$69 \cite{stef08}, makes particularly $^{77}$Cu an interesting physics case. More specifically the lowering of the 5/2$^{-}$ relative to the 3/2$^{-}$ state is expected to be strong enough to have a 5/2$^{-}$ ground state in $^{77}$Cu.  

In the following, we discuss the experimental setup used  (Sec. \ref{exp}) along with the obtained results (Sec. \ref{results}). In the last section the proposed ground-state spin and parity of $^{77}$Cu is discussed. In the same section a short interpretation is given for the spin and parity assignments of the ground state and low-lying excited states of $^{77}$Zn.

\section{Experimental setup} \label{exp}
The neutron-rich copper isotopes were produced via proton- and neutron-induced fission of $^{238}$U target. More specifically the 1 GeV proton beam delivered from the CERN PS-Booster impinged either directly to the thick 50 g/cm$^2$ Uranium carbide target or it was sent to a tantalum proton-to-neutron converter to induce fission through the produced spallation neutrons. For the case of $^{77}$Cu, the latter technique was used as to suppress the Rb isobaric contamination, produced through spallation and present in the mass-separated beam as the surface ionization of this element is not negligible. Another element that is easily surface ionized is Ga. The Ga isotopes are strongly populated by spallation-neutron-induced fission and thus remain present in the mass-separated beam. The $^{77}$Cu nuclei were selectively ionized by means of the Resonant Ionization Laser Ion Source (RILIS) \cite{mis93,fed00,kos03} and mass separated by the General Purpose Separator (GPS) of the ISOLDE facility \cite{kug00}.

\begin{figure}
 \begin{center}
 \includegraphics[scale=0.70] {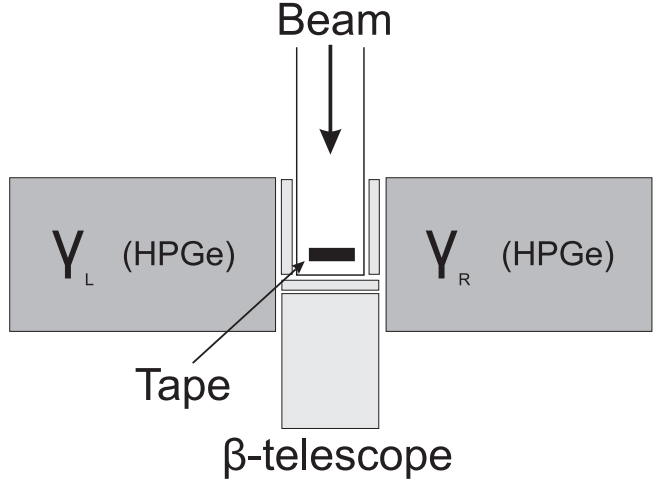}
 \end{center}
 \caption{Top view of the detector geometry at the $\beta$-decay setup. The $\beta$ detectors
are indicated with the light gray colour.
 \label{setup}}
 \end{figure}

After mass separation, the radioactive beam with a typical energy of 60 keV, was delivered to the $\beta$-decay experimental setup (Fig. \ref{setup}). The $^{77}$Cu nuclei were implanted on a tape which was surrounded by three plastic $\Delta$E detectors for the $\beta$-particles detection. The emitted $\gamma$ rays were recorded through two HPGe detectors with relative efficiencies 75\% (left) and 90\% (right). These two detectors were placed in close geometry, typically at 2 cm distance, aiming for the maximum possible detection efficiency. The full energy of the $\beta$ particles was recorded with the $\beta$ telescope which was installed behind the implantation tape at 0$^{\circ}$ with respect to the beam direction.

The electronics read-out was done in parallel by means of two different ways: the singles mode where the energy signals of the two $\gamma$ detectors were recorded together with the E part of the $\beta$ telescope and the List Mode Data (LMD) acquisition mode where the data were recorded in an event by event mode. In the LMD mode, the trigger was created whenever a $\beta-\gamma$, $\beta-\beta$ or a $\gamma-\gamma$ event was recorded. Given the much higher efficiency of the $\beta$ detection setup most of the LMD data corresponds to $\beta$-triggered events. 

For a more precise lifetime determination the $^{77}$Cu beam was sent to a second setup to detect the emitted $\beta$-delayed neutrons. It consisted of the Mainz neutron long counter \cite{arn07} with a built-in tape transport system to remove long-lived background after each measurement cycle. A thin plastic scintillator served for the detection of $\beta$ particles. Despite the selectivity of the RILIS on mass 77 the $\beta$ activity was dominated by gallium. Therefore the $\beta$ rays of the $^{77}$Cu decay could not be discriminated and no $P_n$ value could be deduced. Nevertheless, the $\beta$-delayed neutron signal was very clean since the isobaric contaminants do not emit neutrons. Measurements with the RILIS lasers off or detuned from the resonance gave only a flat background (cosmic radiation).

The timing signals of the neutron counters were fed into an 8-input multichannel-scaler (module 7884 from Fast Comtec) that can handle data rates above 1 MHz. It was read out by a MPA2 DAQ from Fast Comtec.
For calibration, the entire acquisition system was tested at up to two orders of magnitude higher data rates (abundant Rb and Cs isotopes) but showed no significant deadtime effects.

\begin{table}

 \caption{Experimental conditions for the decay study of $^{77}$Cu.}
\label{expCond}
\begin{ruledtabular}
\begin{tabular}{cccc}

 Cycle & Laser on & Laser off \\
 implantation/decay time & measurement time & measurement time \\
  (s) & (min) &  (min)\\
\hline

 0.6s/2.2s & 170 & 115\\
\end{tabular}
\end{ruledtabular}
\end{table}

In Table \ref{expCond}, a summary of the experimental conditions for the decay study of $^{77}$Cu is given. Each run is simply a repetition of cycles, where each cycle has a characteristic implantation and decay period depending on the half life of the isotope under study. After such a cycle, the tape is moved in order to suppress the long-lived background activity. For each isotope, a certain amount of beam time was used without using the lasers (lasers-off runs) as to determine the beam contamination (mainly Ga) and the room background. A more detailed description of the experimental setup can be found in \cite{van05, van02, tho06}

\begin{figure}
\includegraphics[scale=0.4]{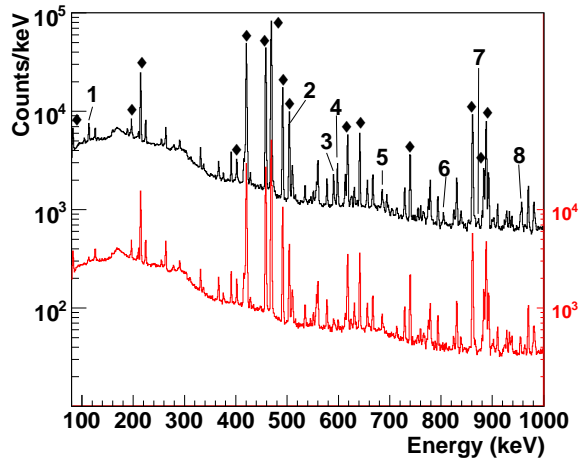}
\caption{\label{spectra}Part of the LMD-based lasers-on (upper black histogram) and lasers-off (lower red histogram) spectra. Lines labelled with a number belong to the $^{77}$Cu decay and are listed in Table \ref{77CuTable1} where the intensity is  given. The transitions indicated with $\blacklozenge$ correspond to the decay of the surface-ionized $^{77}$Ga contaminant.}
\end{figure}

\begin{figure*}
\includegraphics[scale=0.8]{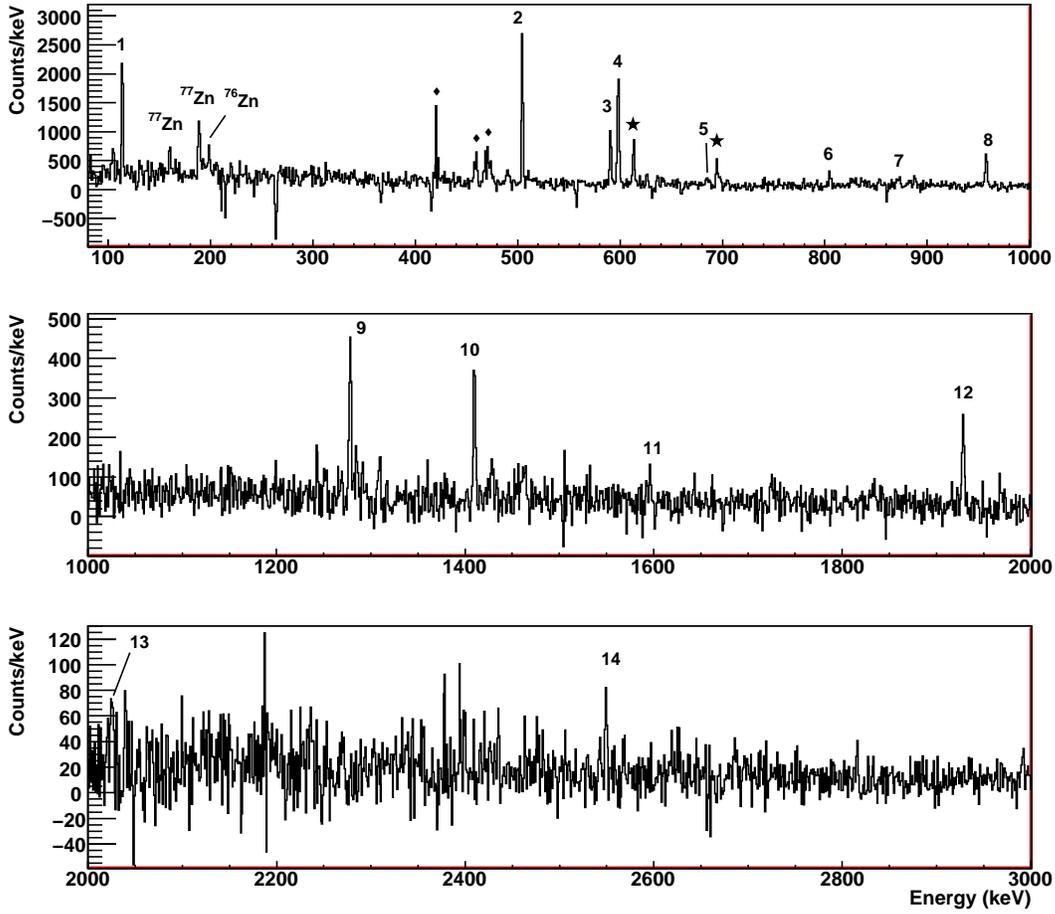}
\caption{\label{spectraDiff} The LMD-based spectrum after the subtraction of lasers-on minus lasers-off spectra shown in Fig. \ref{spectra}. In this subtracted spectrum the activity coming from the $\beta$ decay of $^{77}$Cu can be easily identified (labeled by numbers). The explanation for the negative spikes is given in the text. Also some of the strongest $^{77}$Ga lines ($\blacklozenge$) are present in the subtracted spectrum, together with some long-lived activity of a previous implantation: $^{78}$As ($\star$).}
\end{figure*}

\begin{table*}
\caption{\label{77CuTable1} Energy values and relative intensities of the $^{77}$Cu decay lines in the LMD-based spectrum.
Coincident $\gamma$-rays are given in the last column. Intensities are indicated within brackets. }
\begin{ruledtabular}
\begin{tabular}{cccc}
Label & Energy (keV) & Intensity & Coincident $\gamma$ rays \\
\hline
1 &  114.7 (0.3) & 33 (4) & 685.5(4$\pm$1), 1169.8(6$\pm$1)\\
2 &  505.7 (0.9) & 100 (5) & 638.0(7$\pm$2), 804.5(16$\pm$2), 957.3(40$\pm$3), 1595.3(8$\pm$2), 1662.1(7$\pm$2),\\
 & & &  1805.0(8$\pm$2), 1926.0(33$\pm$2), 2108.0(6$\pm$2), 2549(13$\pm$2) \\
3 & 591.1 (0.2) & 34 (4) & 871.9(10$\pm$2), 1290.8(17$\pm$3), 1840.6(5$\pm$2), 2023.0(13$\pm$2), 2463.6(6$\pm$2)\\
4 & 598.3 (0.3)\footnotemark[1] & 89 (3) & 698.0(16$\pm$2)\\
5 & 685.5 (0.5) & 4 (1) & 114.7(33$\pm$4)\\
6 & 804.5 (1.0) & 16 (2) & 505.7(100$\pm$5), 1277.7(46$\pm$4), 1662.1(7$\pm$2) \\
7 & 871.9 (1.0) & 10 (2) & 591.1(34$\pm$4) \\
8 & 957.3 (0.3) & 40 (3) & 505.7(100$\pm$5), 638.0(7$\pm$2), 1277.7(46$\pm$2)\\
9 & 1277.7 (0.2) & 46 (4) & 957.3(40$\pm$3), 1595.3(8$\pm$2), 1662.1(7$\pm$2), 1805.0(8$\pm$2), 1926.0(33$\pm$2), 2549.0(13$\pm$2)\\ 
10 & 1409.3 (0.3) & 37 (3) & 1463.7(6$\pm$2), 2335.0(7$\pm$3)\\
11 & 1595.3 (1.0) & 8 (2) & 505.7(100$\pm$5)\\
12 & 1926.0 (0.5) & 33 (2) & 505.7(100$\pm$5), 1277.7(46$\pm$4)\\
13 & 2023.0 (1.0) & 13 (2) & 591.1(34$\pm$4)\\
14 & 2549.0 (1.0) & 13 (2) & 505.7(100$\pm$5), 1277.7(46$\pm$4)\\
15 & 3826.7 (1.0) & 3 (1) & - \\
\end{tabular}
\end{ruledtabular}
\footnotetext[1]{$\beta$-delayed neutron emission.}
\end{table*}

\section{Results} \label {results}

The knowledge about the $\beta$-decay properties of $^{77}$Cu is limited. The half life has been determined to be 469 $\pm$ 8 ms in \cite{kra91} and 450$^{+13}_{-21}$ ms in \cite{hos05}. Concerning the levels of the daughter nucleus $^{77}$Zn only a few of them have so far been identified. The 772 keV (1/2$^{-}$ isomeric state, T$_{1/2}$=1.05$\pm$0.10 s) has been identified in \cite{eks86} and the 114.7 keV level together with the 800 keV level have been identified in the $\beta$-delayed neutron emission of $^{78}$Cu \cite{van05}. Only recently, more levels were identified in a decay study at HRIBF \cite{ily07,win09}.

The data set for $^{77}$Cu can be divided into two parts: the data acquired with the lasers-on resonance and the data with the lasers-off resonance. In Fig. \ref{spectra} the main part of the LMD-based data using both HPGe detectors is presented. Both  the lasers-on and lasers-off  spectra correspond to the data recorded during the first 1575 ms after the starting of the implantation period. In this way almost the full statistics of the decay lines from the short-lived $^{77}$Cu decay are obtained while the intensity of the lines corresponding to the long-lived beam contaminants is reduced. As can be seen in Fig. \ref{spectra} the great majority of the transitions observed are non resonant, mainly from the decay of the surface-ionized $^{77}$Ga. The decay scheme of $^{77}$Ga is not very well known \cite{far97} but the identification of the $\gamma$-ray lines from the $\beta$ decay was feasible given its prominent presence in both on-resonance and off-resonance spectra. The most intense lines related to $^{77}$Ga decay are indicated with a diamond symbol ($\blacklozenge$). Other lines that appear both in the lasers-on and lasers-off spectra are  from the $\beta$ decay of long-lived daughter nuclei originating mainly from $^{77}$Ga or from other isotopes deposited in the tape system during previous runs associated with different mass settings.
As can be seen in Fig. \ref{spectraDiff} the laser-enhanced lines can easily be identified after normalizing the two spectra of Fig. \ref{spectra} and subtracting them. The resulted spectrum contains only the peaks related to the $^{77}$Cu decay and daughter activity ($^{77}$Zn). At this point it is worth noting that even in the subtracted spectrum the strongest $^{77}$Ga lines are still present. The negative counts 
are due to long-living $^{77}$Ge (11.3 h) which has been growing in since the
measurements started. Given the fact that the measurement started with the lasers on resonance and then
later with lasers off relatively more counts of $^{77}$Ge are present in the
lasers-off spectrum. The 772 keV transition is not present in the coincidence spectra due to the long half-life time ($\sim$ 1s \cite{eks86}) of the isomeric state. Since the decay of $^{77}$Zn is well known \cite{far97} the identification of resonant lines attributed to the $^{77}$Cu decay or to its daughter $^{77}$Zn was feasible. In Figs. \ref{spectra} and \ref{spectraDiff} the lines assigned to the $\beta$ decay of $^{77}$Cu are labelled  with a number.





\begin{figure}
 \begin{center}
 \includegraphics[scale=0.4] {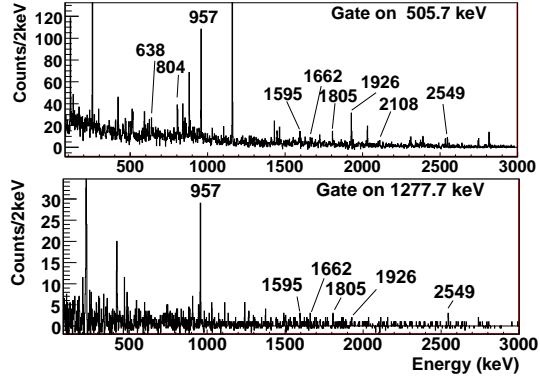}
 \end{center}
 \caption{The $\gamma$-ray spectrum of prompt coincidence events with
 506 keV (upper panel) and 1278 keV (lower panel) lines. Only the lines from $^{77}$Cu decay are labelled. Unlabelled lines come from $^{77}$Ga contamination.  
 \label{coinc}}
 \end{figure}

Information on the observed $^{77}$Cu decay lines is summarized in Table \ref{77CuTable1}. The rather high uncertainty on a number of intensities reflects the fact that many of the lines were contaminated. In these cases, the uncertainty of the resulted peak area is much higher than in the cases of single isolated peaks - free of contaminants (e.g. the 1409.3 keV line). As a second step of control of the $\gamma$-ray assignment procedure, the net area of each transition in the subtracted spectrum was determined by setting different conditions on the absolute time of each coincident event with respect to the proton pulse of the primary beam. In this way for all the cases reported in Table \ref{77CuTable1} the time evolution of the net area counts was determined and was found to be compatible with the expected half-life of $^{77}$Cu. The coincidence $\gamma$ rays were derived from background-subtracted $\gamma$-ray gated spectra with a prompt time gate of 0.6 $\mu$s. In Fig. \ref{coinc}, two coincident $\gamma$-ray
spectra are presented. As can be seen in both spectra, several lines are coming from the $^{77}$Ga activity - especially 
for the 505.7 keV case which is a doublet. 
These background lines could be identified since they are also present in the laser-off spectra generated with the same gate conditions.

\begin{figure}
 \begin{center}
 \includegraphics[scale=0.3] {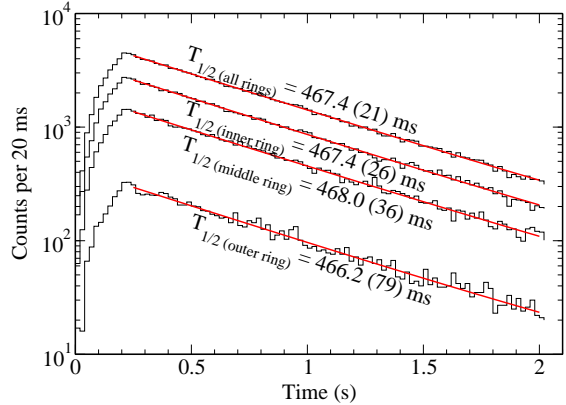}
 \end{center}
 \caption{The counting rate of the Mainz neutron long counter during the implantation/decay cycle. 
 The total rate together with the individual rate of each ring are presented with the corresponding exponential fits.   
 \label{neutrons}}
 \end{figure}

Due to the fact that most of the lines were contaminated with other lines, an accurate half-life could only be deduced from the time behavior of two lines: the 957 keV and the 1409 keV line. The resulted half-life (T$_{1/2}$=467$\pm$43 ms) is less precise than but consistent with the reported values in \cite{kra91,hos05} and with our value out of the $\beta$-delayed neutron measurement. As can be seen in Fig. \ref{neutrons}, a single exponential fit of the neutron-counting rate gives a half-life of 467.4(21) ms for $^{77}$Cu, consistent with, but far more precise than the published values \cite{kra91,hos05}. The latter value will be further used.

The Mainz neutron long counter consists of three concentric rings of $^3$He tubes. Low-energy neutrons are most likely detected in the innermost ring, higher-energy neutrons cover a larger distance before being moderated and detected and are therefore more likely detected in the middle or outer ring. The so-called ``ring ratios'', i.e. the relative rate of neutrons detected in the middle to inner or outer to inner ring allow to conclude on the
{\em average} energy of the $\beta$-delayed neutrons \cite{ree81}. The energy-dependent detection efficiency was simulated with the Monte Carlo neutron transport code MCNP. The simulations were validated with well-known emitters of $\beta$-delayed neutrons.
Thus, from the ring ratio we found an average neutron energy of about 450(150) keV for $^{77}$Cu decay.

With a Q$_{\beta}$ value of 10.1(4) MeV for $^{77}$Cu from the Audi-Wapstra systematics \cite{aud03}, a neutron binding energy of 4.557 MeV for $^{77}$Zn \cite{hak08}, and apparent feedings of 21.6\% and 4.7\% respectively to the 2$_1^+$ and 4$_1^+$ states in $^{76}$Zn we deduce apparent log ft values of 4.1(2) and 4.4(2) to the neutron emitting states populating by neutron emission the 2$_1^+$ and 4$_1^+$ states in $^{76}$Zn. Note that frequently several neutron emitting levels populate the same level in the $A-1$ daughter. Therefore the given log ft values are {\it cumulative} over all neutron emitting states populating the 2$_1^+$ and 4$_1^+$ states in $^{76}$Zn. The uncertainty of the log ft values is dominated by the uncertainty in the $Q_\beta$ value of $^{77}$Cu.

\begin{figure*}
 \begin{center}
 \includegraphics[scale=0.70] {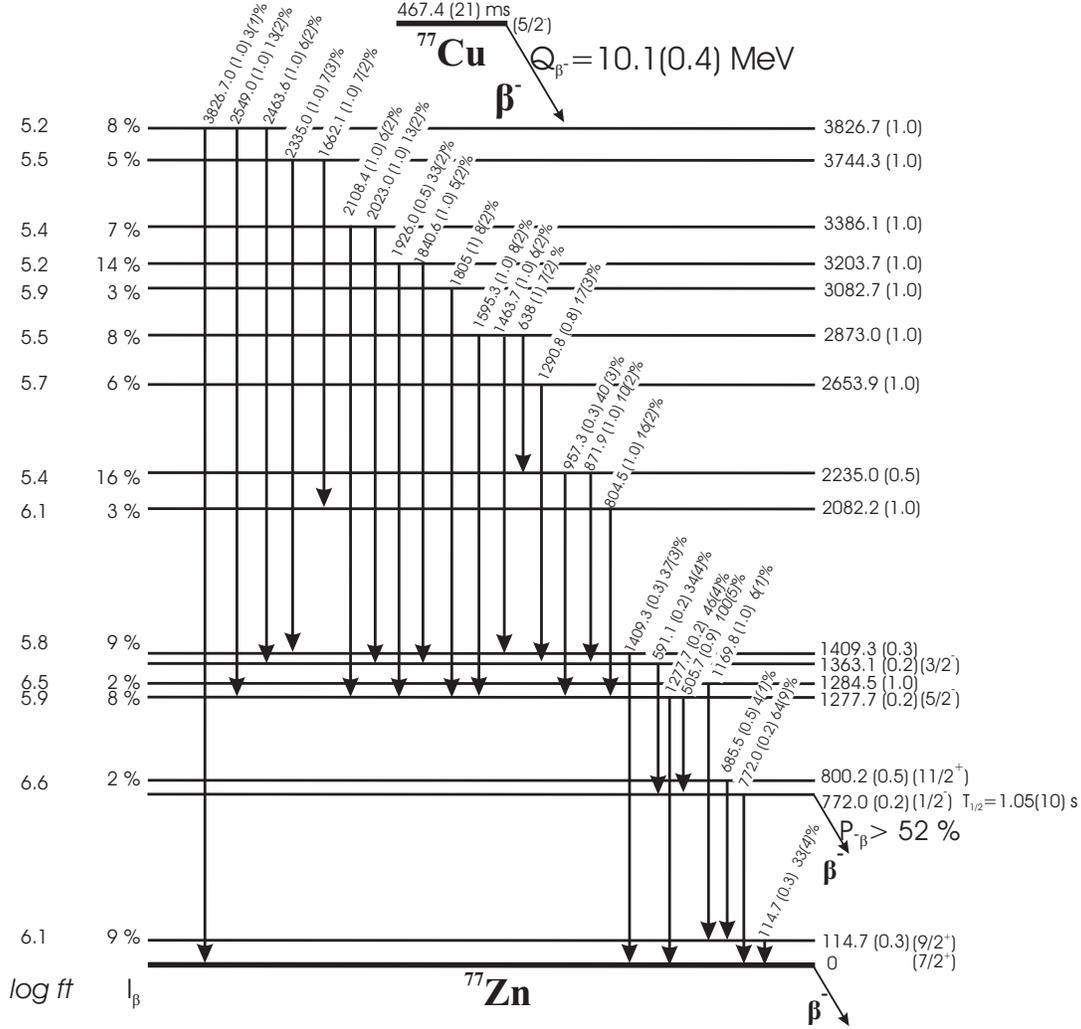}
 \end{center}
 \caption{Deduced decay scheme for $^{77}$Cu. The Q$_{\beta^{-}}$ value was taken from \cite{aud03}. The proposed spin assignments are discussed in the text. The given $\beta$-branching values should be considered only as upper limits and the \textit{log ft} values as lower limits.
 \label{decay77}}
 \end{figure*}

\begin{figure*}
 \begin{center}
 \includegraphics[scale=0.60] {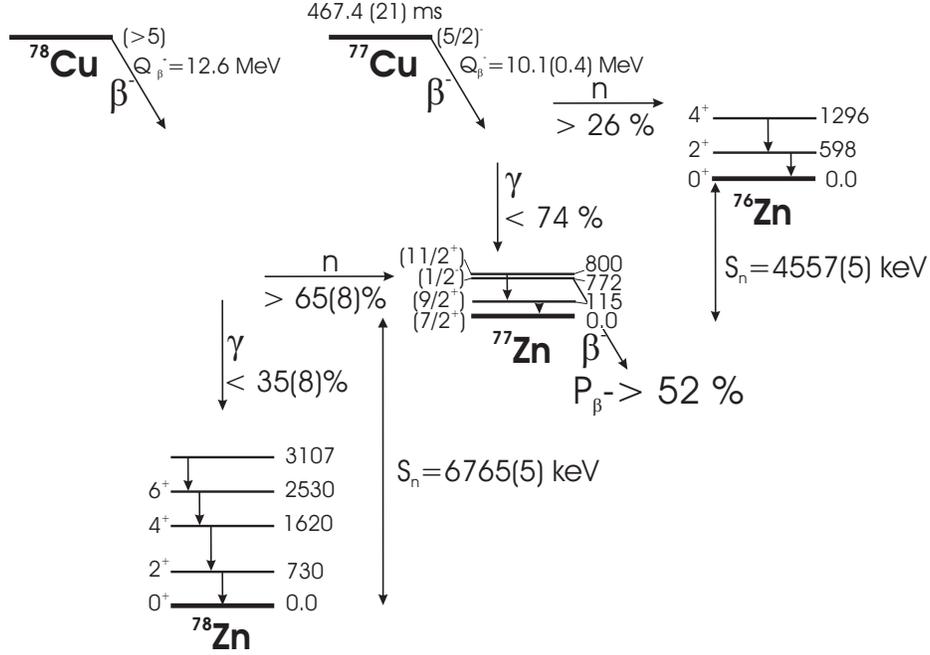}
 \end{center}
 \caption{Abbreviated decay scheme of $^{78,77}$Cu with emphasis on the $\beta$-delayed neutron emission branch. Information on the $\beta$-delayed n decay of $^{78}$Cu is taken from \cite{van05} and \cite{win09}. Neutron binding energies are derived from \cite{hak08}.
 \label{neutronEm}}
 \end{figure*}

The decay scheme for $^{77}$Cu shown in Fig. \ref{decay77} could be constructed based on the coincidence relations given in Table \ref{77CuTable1}.
Furthermore, use has been made of the $\beta$-delayed neutron-decay data from $^{78}$Cu \cite{van05} (see also Fig. \ref{neutronEm}). In this decay, $\gamma$ rays of 115 keV and 686 keV were seen in cascade and are also observed here in the $\beta$-decay of $^{77}$Cu. The strong 115 keV line is only coincident with the 686 and 1170 keV lines. The three lines have no connection with other $^{77}$Cu decay transitions. In principle, as long as no connecting transitions are found, the 115 keV line could be feeding the (1/2$^-$) isomeric level at 772 keV instead of the ground state as proposed in \cite{van05} and also assumed here. The present level scheme is in general agreement with the partial level scheme in \cite{ily07,win09} but extends it considerably.

The relative $\gamma$-ray intensities were calculated from the LMD-based data taking into account the variation of the detection efficiency with respect to the $\gamma$-ray energy. Where necessary, the lines were corrected for contaminating $\gamma$ rays using the off-resonance data. The relative intensities were calculated with respect to the most intense line at 505.7 keV and are given in Table \ref{77CuTable1}. The close detection geometry can, depending on the multiplicity of the $\gamma$-decay cascade, lead to summing distortions in the intensities. As only relative $\gamma$-ray intensities are given and the quoted intensity uncertainties are rather high, these distortions can be neglected. Concerning the isomeric 772.0 keV line, the observed counts were corrected for the mother-daughter decay relation taking into account the implantation and decay cycle.  For those transitions observed only in the $\gamma-\gamma$ coincidence spectra, coincidence relations and comparison with other lines were used to extract the intensities.

The $\beta$ feeding shown in Fig. \ref{decay77} was estimated from the intensities given in Table \ref{77CuTable1} assuming the feeding to the ground state and to the isomeric state at 772 keV equal to zero. The same was assumed for the 1363 keV level since the intensity of the 591 keV $\gamma$ line de-exciting this level is within the error-bars equal to the total intensity of the $\gamma$ rays that feed this level. The experimental determination of the possible existence of $\beta$ feeding to the ground state and to the isomeric state was impossible since no singles $\beta$ spectrum was acquired that could be compared with the $\beta-\gamma$ coincidence spectrum. Therefore, and also due to unobserved $\gamma$-ray feeding,  the given $\beta$-branching values should be considered as upper limits and the \textit{log ft} values as lower limits. The quoted Q$_{\beta}$ value was taken from \cite{aud03}. Based on the assumption that there is no direct $\beta$ feeding to the (1/2$^-$) state at 772 keV, a lower limit of 52\% for the $\beta$-decay probability of this state can be deduced on the basis of the balance of feeding and de-excitation $\gamma$-ray intensities. A remarkable feature of the $^{77}$Cu decay scheme is the fact that the $\gamma$ rays de-exciting the high-energy levels are feeding a limited number of excited states avoiding intense feeding of the (1/2$^{-}$) isomeric state or the (7/2$^{+}$) ground state. A similar pattern is observed in the decay of $^{75}$Cu and $^{73}$Cu \cite{fau08}.

In Fig. \ref{neutronEm}, a partial decay scheme of $^{78,77}$Cu is presented with emphasis on the $\beta$-delayed neutron-emission branch. The strong feeding of the 4$^+$ and especially 2$^+$ level in $^{76}$Zn in the $\beta$-delayed neutron-emission of $^{77}$Cu, indicates that one or more $^{77}$Zn high-energy levels above the neutron binding energy get a large fraction of the overall $\beta$ branching. From the intensity balance of the 598.3 keV $\gamma$ ray and the $^{77}$Zn $\gamma$ rays, a lower limit for the probability for neutron emission in $^{77}$Cu was obtained: P$_{n}\geq$26\% in agreement with the recent results of Winger {\it et al} \cite{win09} and much higher than reported in \cite{pfe02}. 
Taking into account the spin and energy selection rules for $\beta$ decay and neutron emission and the fact that the $\beta$-delayed neutron channel relatively populates for 18\% the 4$^+$ state and for 82\% the 2$^+$ state in $^{76}$Zn limits the possible spin for the $^{77}$Cu ground state to 5/2 and 7/2. Combining this information with the fact that the $\beta$-delayed $\gamma$ emission of $^{77}$Cu does feed levels that finally ends up for relatively 54\% on the (1/2$^-$) isomeric state and for 46\% on the (7/2$^+$) ground state favours a 5/2$^-$ assignment for the ground state of $^{77}$Cu. In this relative feeding, the 772 keV isomeric transition from the (1/2$^-$) to the (7/2$^+$) is not taken into account. The proposition for a 5/2$^-$ spin and parity of $^{77}$Cu is in agreement with recent results from laser spectroscopy experiments at ISOLDE where a 3/2$^-$ assignement could be excluded leaving only 5/2 or 7/2 as possible spins \cite{fla08}. 
In reference \cite{ily07}, the $\beta$-decay study of $^{78}$Cu seems to point to a different feeding pattern than observed in our study of $^{78}$Cu \cite{van05} where the 910 keV 6$^{+} \to$ 4$^+$ transition in $^{78}$Zn was not seen. Reanalysis of the data confirms the decay scheme of \cite{ily07} (see also Fig. \ref{neutronEm}) through the observation of the 576 keV line de-exciting the 3107 keV level with an intensity of 17$\pm$10 \% relative to the 730 keV 2$^{+}\to$ 0$^+$ transition and a weak peak at 910 keV line (6$^{+} \to$ 4$^+$ transition) with a relative intensity of 12$\pm$9 \%. The observation of the 6$^+$ level, fed in the decay of $^{78}$Cu, suggests a spin of at least 5 for the ground state of $^{78}$Cu. This high spin makes it unlikely that the $\gamma$ transitions observed in the $\beta$-delayed neutron emission of $^{78}$Cu (115 keV, 686 keV) can be placed on top of the (1/2$^-$) isomer in $^{77}$Zn but rather are connected to the ground state (7/2$^+$). The spin and parity suggestion of (9/2$^+$), (11/2$^+$) of respectively the 115 keV, 800 keV levels is based on the feeding pattern observed in the $^{77,78}$Cu decay. The present decay scheme of $^{77}$Cu is certainly not complete as the relative strong population of the low-lying positive parity states cannot result from direct $\beta$ feeding from the (5/2$^-$) $^{77}$Cu ground state. Two explanations are possible: missed $\gamma$-ray feeding from higher-lying levels or the existence of a $\beta$-decaying isomer in $^{77}$Cu. The latter can propably be excluded as there is no evidence seen in the laser spectroscopy studies of $^{77}$Cu \cite{fla08}.

\section{Discussion} \label {Discussion}
\begin{figure}
 \begin{center}
 \includegraphics[scale=0.40] {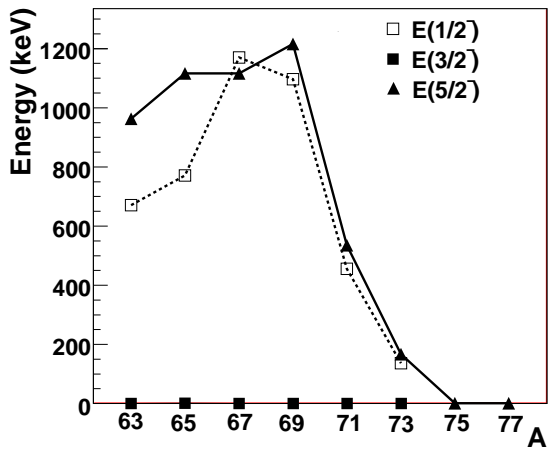}
 \end{center}
 \caption{Systematics of the energies of the first excited 1/2$^-$, 5/2$^-$ and 3/2$^-$in $^{63-77}$Cu (based on \cite{stef08,fra98,fla08} and references therein).
 \label{levels}}
 \end{figure}

According to the results of the present work, a 5/2$^-$ spin assignment is proposed for the ground state of $^{77}$Cu. In  Fig. \ref{levels}, the systematics of the 1/2$^-$, 5/2$^-$ and 3/2$^-$ states which correspond to the ground and first excited states of the neutron-rich odd Cu isotopes are presented. As can be seen in this figure  for $^{63-69}$Cu, the position of the first 5/2$^-$ level lies almost constantly 1 MeV above the 3/2$^-$ ground state. However, when the $\nu$g$_{9/2}$ neutron orbital starts to be filled, the gap between those levels is dramatically decreased. From the experimental determination of the B(E2; 5/2$^{-}$ $\to$ 3/2$^{-}$) values it has become clear that the nature of the first 5/2$^{-}$ state in the odd Cu isotopes changes from a core-coupled state in the N$<$40 isotopes to a single particle state in the N$\ge$40 isotopes.  The single particle character of the 5/2$^-$ state is  established from $^{69}$Cu onwards \cite{stef08}.  The lowering of the 5/2$^-$ state is strong enough to become the ground state in $^{75}$Cu \cite{fla08}. The observed sharp drop in energy of the 5/2$^-$ level, known also as monopole migration, is believed to be caused by the attractive proton-neutron j$_{>}$-j$_{<}$ tensor force \cite{ots05} (see also \cite{stef08,fra01}).

The low-energy level structure of the neutron-rich odd-mass Zn nuclei is dominated by the p$_{3/2}$, f$_{5/2}$, p$_{1/2}$ and g$_{9/2}$ neutron orbitals. The gradual lowering in energy of the unique parity state g$_{9/2}$ as a function of increasing neutron number leads from $^{69}$Zn on to isomerism. In fact, from there on the states based on the p$_{1/2}$ and on the g$_{9/2}$ orbitals lie within 500 keV of each other \cite{fau08}. The situation is different in $^{77}$Zn as the (1/2$^-$) state has risen up to 772 keV above the (7/2$^+$) ground state. The positive parity levels assigned as 7/2$^+$ and 9/2$^+$ can be produced by quasi-particle couplings of the ($\nu$g$_{9/2}$)$^{-3}$ configuration. A similar system where three quasi-particles in $\pi$g$_{9/2}$ orbit lead to a low-lying 7/2$^+$ state has been also observed in the Ag isotopes where three holes in the $\pi$g$_{9/2}$ shell result in a similar ordering of the 7/2$^+$ and 9/2$^+$ levels. The 1/2$^-$ isomeric state can be interpreted as a $\nu$p$_{1/2}$ single particle state. 
Coulomb-excitation and laser-spectroscopy studies for the odd-mass Zn isotopes that can be performed in the future as well as the results of the recently performed laser spectroscopy experiments for the neutron rich Cu isotopes \cite{fla08} could contribute for a better understanding of the residual proton-neutron interaction in this mass region and the role of the 1g$_{9/2}$ orbital.

\begin{acknowledgements}
This work was supported by he IUAP - Belgian Science Policy - BriX network (nr. P6/23), FWO Vlaanderen (Belgium) and GOA/2004/03 (BOF-K.U.Leuven). The authors would like to thank B. Walters for stimulating discussions and K.L. Kratz for the use of Mainz neutron long counter.
\end{acknowledgements}


\begin{thebibliography}{100}




\bibitem{sor08} O. Sorlin, M.-G. Porquet, Progress in Particle and Nuclear Physics
{\bf 61}, 602 (2008).
\bibitem{bre08} N. Bree {\it et al}., Phys. Rev. C {\bf 78}, 047301 (2008).
\bibitem{sor02} O. Sorlin {\it et al}., Phys. Rev. Lett. {\bf 88}, 092501 (2002).
\bibitem{bro95} R. Broda  {\it et al}., Phys. Rev. Lett. {\bf 74}, 868 (1995).
\bibitem{gra01} H. Grawe and M. Lewitowicz, Nucl. Phys. A {\bf693}, 116 (2001).

\bibitem{lan03} K. Langanke, J. Terasaki, F. Nowacki, D. J. Dean, and W. Nazarewicz, Phys. Rev. C {\bf 67}, 044314 (2003).

\bibitem{gue07} C. Gu{\'e}naut {\it et al}., Phys. Rev. C {\bf 75}, 044303 (2007).
\bibitem{rah07} S. Rahaman {\it et al}., Eur. Phys. J. A {\bf 34}, 5 (2007).


\bibitem{stef08} I. Stefanescu {\it et al}., Phys. Rev. Lett. {\bf 100}, 112502 (2008).
\bibitem{stef07} I. Stefanescu {\it et al}., Phys. Rev. Lett. {\bf 98}, 122701 (2007).
\bibitem{van07} J. Van de Walle {\it et al}., Phys. Rev. Lett. {\bf 99}, 142501 (2007).
\bibitem{van09} J. Van de Walle {\it et al}., Phys. Rev. C {\bf 79}, 014309 (2009).
\bibitem{mue99} W.F. Mueller {\it et al}., Phys. Rev. Lett. {\bf 83}, 3613 (1999).
\bibitem{fra98} S. Franchoo {\it et al}., Phys. Rev. Lett. {\bf 81}, 3100 (1998).
\bibitem{van05} J. Van Roosbroeck {\it et al}., Phys. Rev. C {\bf 71}, 054307 (2005).
\bibitem{kra93} K.-L. Kratz {\it et al}., Astrophys. J. {\bf 403}, 216 (1993).
\bibitem{kug00} E. Kugler {\it et al}., Hyp. int. {\bf 129}, 23 (2000).
\bibitem{van02} J. Van Roosbroeck, PhD thesis, IKS-K.U.Leuven (2002).
\bibitem{tho06} J.-C. Thomas {\it et al}., Phys. Rev. C {\bf 74}, 054309 (2006).

\bibitem{van204} J. Van Roosbroeck {\it et al}., Phys. Rev. Lett. {\bf 92}, 112501 (2004).
\bibitem{van04} J. Van Roosbroeck {\it et al}., Phys. Rev. C {\bf 69}, 034313 (2004).
\bibitem{fau08} T. Faul, PhD thesis, IReS, Strasbourg (2008).
\bibitem{mis93} V. Mishin {\it et al}., Nucl. Instr. and Meth. {\bf 73}, 550 (1993).
\bibitem{fed00} V. Fedoseyev {\it et al}., Hyp. Inter. {\bf 127}, 109 (2000).
\bibitem{kos03} U. K{\"o}ster {\it et al}., Spectrochimica Acta B. {\bf 58}, 1047 (2003).
\bibitem{arn07} Oliver Arndt, Ph.D. thesis,
Johannes-Gutenberg-Universit\"at Mainz (2007).

\bibitem{kra91} K.-L. Kratz {\it et al}.,Z. Phys. A {\bf 340}, 419 (1991).
\bibitem{hos05} P.T. Hosmer {\it et al}., Phys. Rev. Lett. {\bf 94}, 112501 (2005).
\bibitem{eks86} B. Ekstr{\"o}m {\it et al}.,Physica Scr. {\bf 34}, 614 (1991).
\bibitem{ily07} S.V. Ilyushkin {\it et al}, Proceedings of the 4th International Conference on Fission and Properties of Neutron-Rich Nuclei (2007).
\bibitem{win09} J.A. Winger {\it et al}., Phys. Rev. Lett. {\bf 102}, 142502 (2009).
\bibitem{far97} A.R. Farhan and B. Singh, Nuclear Data Sheets {\bf 81}, 417 (1997).
\bibitem{ree81} P.L. Reeder and R.A. Warner,  Nucl. Instr. and Meth. {\bf 180}, 173 (1981).
\bibitem{aud03} G. Audi, A.-H. Wapstra, and C. Thibault, Nucl. Phys. {\bf A729}, 337 (2009).
\bibitem{hak08} J. Hakala {\it et al}., Phys. Rev. Lett. {\bf 101}, 052502 (2008).
\bibitem{pfe02} B. Pfeiffer and K.-L.Kratz, Prog. Nucl. Energy {\bf 41}, 39 (2002).

\bibitem{fla08} K.T. Flanagan, private communication

\bibitem{ots05} T. Otsuka, T. Suzuki, R. Fujimoto, H. Grawe, and Y. Akaishi, Phys. Rev. Lett. {\bf 95}, 232502 (2005).
\bibitem{fra01} S. Franchoo {\it et al}., Phys. Rev. C {\bf 64}, 054308 (2001).


\end{thebibliography}
\end{document}